\documentclass[pdftex,twocolumn]{aastex6}

\usepackage{apjfonts,graphicx,graphics}

\usepackage{graphicx}
\usepackage{natbib}
\usepackage{mathptmx}
\usepackage{amsmath}
\usepackage{wasysym}
\usepackage{times}
\usepackage{epsfig}
\usepackage{ulem}

\begin{document}

\title{Constraint on cosmological parameters by Hubble parameter from gravitational wave standard sirens of neutron star binary system}
\author{Liu Yang$^{1}$, Hao-Yi Wan$^{1,2}$, Tong-Jie Zhang$^{1,3}$, Tang Yanke$^{3}$}
\affil{$^1$ Department of Astronomy,Beijing Normal University Beijing 100875, China; tjzhang@bnu.edu.cn\\
       $^2$ National Astronomical Observatories, Chinese Academy of Sciences, Beijing 100012, China\\
       $^3$ College of Physics and Electronic information, Dezhou University, Dezhou 253023, China.}

\begin{abstract}

In this paper, we present a new method of measuring Hubble parameter($H(z)$), making use of the anisotropy of luminosity distance($d_{L}$), and the analysis of gravitational wave(GW) of neutron star(NS) binary system. The method has never been put into practice before due to the lack of the ability of detecting GW. LIGO's success in detecting GW of black hole(BH) binary system merger announced the possibility of this new method. We apply this method to several GW detecting projects, including Advanced LIGO(Adv-LIGO), Einstein Telescope(ET) and DECIGO, finding that the $H(z)$ by Adv-LIGO and ET is of bad accuracy, while the $H(z)$ by DECIGO shows a good accuracy. We use the error information of $H(z)$ by DECIGO to simulate $H(z)$ data at every 0.1 redshift span, and put the mock data into the forecasting of cosmological parameters. Compared with the available 38 observed $H(z)$ data(OHD), mock data shows an obviously tighter constraint on cosmological parameters, and a concomitantly higher value of Figure of Merit(FoM). For a 3-year-observation by standard sirens of DECIGO, the FoM value is as high as 834.9. If a 10-year-observation is launched, the FoM could reach 2783.1. For comparison, the FoM of 38 actual observed $H(z)$ data is 9.3. These improvement indicates that the new method has great potential in further cosmological constraints.

\end{abstract}


\section{Introduction}
\label{sec:intro}

In the twenty-first century, we witnessed the bloom of accurate cosmology. Accurate cosmology even ranked second on a list named "Insights of the decade" from Science magazine in 2010. The key of accurate cosmology is to accurately constrain cosmological parameters and their state equations, which can lead us to a better understanding of the evolution of our universe. We mainly developed four observations to constrain cosmological parameters: Supernova(SN), Baryon Acoustic Oscillation(BAO), Galaxy Cluster(CL), Weak Lens(WL) so far(\citealt{2006astro.ph..9591A}). Actually, a relatively new tool, Hubble parameter( $H(z)$ ), is becoming increasingly popular these years because of its effective constraint on cosmological parameters. $H(z)$'s high efficiency lies on the fact that it is the only observation that can directly represent the expanding history of our universe. Compared with the Luminosity distance($d_{L}$) of SN, $H(z)$ contains no integral terms and directly connects with cosmic parameters, which makes it a powerful tool in constraining cosmological parameters, because the integral term can conceal many details and hide important information. As Ma and and Zhang reported, $H(z)$ constrains cosmological parameter much tighter than the same-number SN does. To achieve the same constraint effect of $H(z)$, ones need four times as many SNs as $H(z)$(\citealt{2011ApJ...730...74M}). Getting a more accurate measurement of $H(z)$ could encourage the development of accurate cosmology a lot.\\

There are various ways to detect $H(z)$, which can be generally classified into three types: 1, differential age method(\citealt{2010JCAP...02..008S}); 2, radial BAO method(\citealt{2009MNRAS.399.1663G}); 3, standard sirens method(\citealt{2009PhRvL.103i1302G}). The first two techniques have been employed in the past detection of $H(z)$, but the data of observed hubble parameter data(OHD) are still insufficient. We get only 38 OHDs so far, whose accuracies are far from desirable. Now with the development of GW detecting technology, it is time to look forwards to the third method: GW standard sirens. In 2015, even the second generation GW detector Adv-LIGO operated not at its design sensitivity, it still detected the first GW signal at its first run(\citealt{2016PhRvL.116f1102A}).  According to theoretical understanding, formula of GW of binary system encodes the information of luminosity $d_{L}$, providing an access to the direct measurement of $d_{L}$. Several frequency windows of GW are targeted by different detectors. The second generation detector are mainly aimed at frequency window $10-1000 Hz$, such as LIGO and VIRGO. The next generation detector plan to reach lower frequency region. The underground project DECIGO was designed most sensitive at $0.1-10 Hz$, while the Einstein Telescope(ET) may also reach the frequency of the order $1 Hz$. The space-based eLISA can even detect GW of $10^{-4}-10^{-1} Hz$. In this paper, we make use of GW sirens to measure $H(z)$ by estimating the error of $d_{L}$. Because NS binary system are used as the source of GW in this paper, the frequency of the GW signal of whom mostly concentrate on $10-1000 Hz$, we ignore the projects whose optimal sensitivity are far away from $10-1000 Hz$, such eLISA, and choose the ones whose optimal sensitivity locate around $10-1000 Hz$.

In 2006, a new way to narrow the relatively error of $H(z)$ by studying the dipole of $d_{L}$ has been proposed(\citealt{2006PhRvL..96s1302B}). But the problem is that the new method needs plenty of SNs if we want to get a relatively accurate $H(z)$, which can not be met in reality. And this method has problem in detecting high-$z$ $H(z)$. Still, it is an instructive idea. Atsushi Nishizawa and Atsushi Tamga et al. gave us an alternative by pointing it out that we can get the error information of $d_{L}$ through the gravitational wave function of NS binary system, instead of SN (\citealt{2011PhRvD..83h4045N}).  As the technology developing, now Adv-LIGO already detected gravitational wave, which was really a big strick. It not only proved the general relativity, but also implied that the new method to measure $H(z)$ was feasible. We sense the possibility and potential from detecting gravitational wave. It is meaningful and cutting-edge to study GW detecting projects. In this paper, we focus mainly on two aspect: 1, How will it work out if we apply the new method to some other projects? 2, with the new $H(z)$ information we get, to what degree could we constrain cosmological parameters?\\

This paper is organized as follows. In sec 2, we are going to sketch the idea of GW standard sirens method by Atsushi Nishizawa \emph{et al.}(\citealt{2011PhRvD..83h4045N}), and apply it to some GW detecting projects. In sec 3, we simulate the $H(z)$ data, and analyze the constraining ability of the mock data. In sec 4, we discuss the result. All through this paper, we adopt the natural unit, $c=G=1$.\\

\section{Method}

\subsection{dipole of luminosity distance}

\begin{figure}[htb]
\center{\includegraphics[width=9cm]  {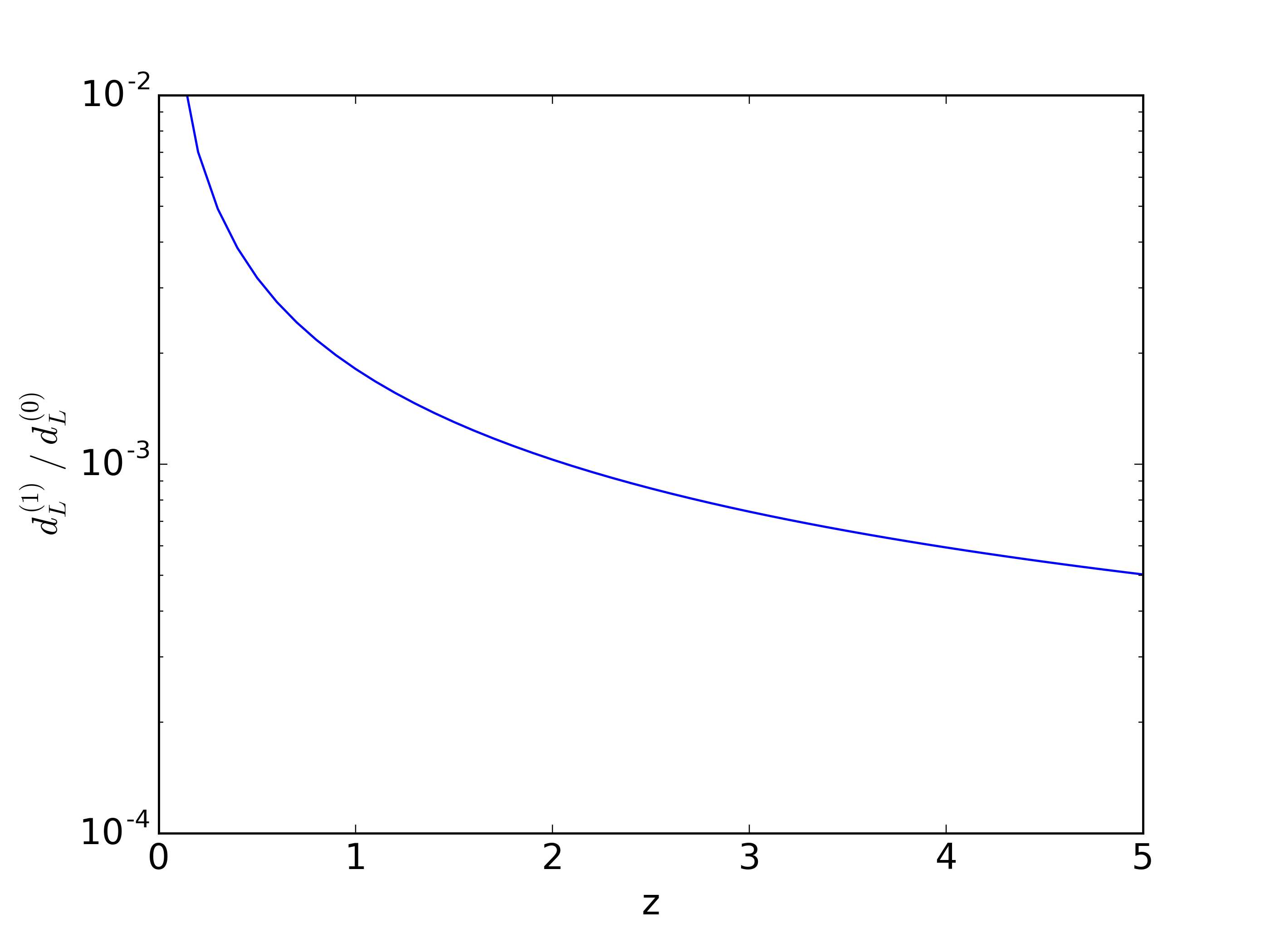}}
\caption{\label{1} The value of $d_{L}^{(1)}/d_{L}^{(0)}$ at different redshift, for $v_{0}=369 km/s$(\citealt{2011ApJS..192...14J}). As is shown in the picture, the ratio goes very large ,even bloom up, at low redshift. That is caused by the ratio approximate to $(1+z)|v_{0}|/z$ at the limit of $z=0$. But this kind of z is not the area we concentrate on.}
\end{figure}

If the universe is completely homogeneous and isotropic on large scale, and the observer is relatively rest with the cosmic microwave background(CMB), the luminosity distance, $d_{L}$, would be just the same form and expression as in standard cosmology. But in fact, there are perturbations around ideal condition leading into the appearance of correction term, multiple of $d_{L}$ (\citealt{1987MNRAS.228..653S}).Then $d_{L}$ can be written as follow:
   \begin{center}
    $d_{L}=d_{L}^{(0)}+d_{L}^{(1)}+$higher order terms
   \end{center}
$d_{L}^{(0)}$represents the traditional meaning of luminosity distance in unperturbed Friedmann universe , also the average of $d_{L}$ on all direction. $d_{L}^{(1)}$ means the dipole of $d_{L}$. As to the "higher order terms", it is self-explantory. The contribution to "higher order terms" coming from the weak gravitational lens effect is so small when compared with dipole that we are going to ignore them here(\citealt{2006PhRvD..73b3523B}). The dipole is dominated by the peculiar velocity of observers. If you want to check it or feel intrigued by the theory, you can look up reference[\citealt{2006PhRvL..96s1302B}] for the details. Here is the final result:
\begin{equation}
\left\{
\begin{aligned}
               & d_{L}^{(1)}=\dfrac{(1+z)^{2}}{H(z)} |v_{0}|\\
               &\dfrac{\Delta H(z)}{H(z)}=\sqrt{3} \left[ \dfrac{d_{L}^{(1)}}{d_{L}^{(0)}} \right]^{-1} \left[ \dfrac{\Delta d_{L}^{(0)}}{d_{L}^{(0)}} \right]   \\
\end{aligned}
\right.
\end{equation}
where $|v_{0}|$, z, $H(z)$ respectively denotes the projection of observer peculiar velocity on the direction of sight, the redshift of the observed celestial body, the expanding rate at the redshift z, and $\Delta d_{L}^{(0)}$,$\Delta d_{L}^{(1)}$ means the error of $d_{L}^{(0)}$,$d_{L}^{(1)}$. The mean error of $H(z)$ will reduce to $\Delta H(z)/ \sqrt{N}$ if we observe N independent sources at the given redshift. Thus, we can improve the accuracy of $H(z)$ by the observation of a large number of sources. From the equations above, given the value of $d_{L}^{(1)}/ d_{L}^{(0)}$ and $\Delta d_{L}^{(0)}/d_{L}^{(0)}$, $\Delta H(z)/H(z)$ can be easily calculated by multiplication. We already know the meaning and expression of $d_{L}^{(0)}$ and $d_{L}^{(1)}$. The result of the term $d_{L}^{(1)}/d_{L}^{(0)}$ is shown in Fig. \ref{1}. To get $\Delta H(z)/H(z)$, the only remaining problem is to find out $\Delta d_{L}^{(0)}/d_{L}^{(0)}$, which can be solved by analyzing observed GW function in following subsection.\\

\subsection{GW standard siren}

One can use SN to illustrate the method of reducing the error of $H(z)$. But due to the number and distribution of SN, it works not that well, especially at high-z region. Considering the advantage of the larger number of observed sources, which can dramatically narrow down the error of $H(z)$, we choose NS binary system as an alternative of SN. What's more, the distribution and property of NS binary system is easier to estimate than that of black hole binary system. We are more familiar with NS. And for black hole binary system, there is a serious problem: black hole seldom radiates electromagnetic wave, rendering it impossible to measure its corresponding redshift up to now. This is an important factor to choose NS binary system \\

In GW experiments, one can extract the property of the source and cosmological information by comparing detected waveform with theoretical template.  That is exactly what LIGO team did when the first detected the GW of two back holes merged(\citealt{2016PhRvL.116f1102A}). The typical Fourier transform of GW waveform can be expressed by the following formula of frequency:
\begin{center}
    $\widetilde{h}(f)=\dfrac{A}{d_{L}(z)}M_{z}^{5/6}f^{-7/6}e^{i\Psi(f)}$
\end{center}
$A=(\sqrt{6}\pi^{2/3})^{-1}$ is a constant which is already geometrically averaged over the inclination angle of a binary system. $d_{L}(z)$ is the luminosity distance at redshift z, and we can set it as $d_{L}^{(0)}$ cause we are going to observe plenty of source at the given redshift. $M_{z}=(1+z)\eta^{3/5}M_{t}$ with the definition of total mass $M_{t}=m_{1}+m_{2}$ and symmetric mass ratio $\eta=m_{1}m_{2}/M_{t}^{2}$. The last unknown function $\Psi(f)$ is a little intricate. It is the frequency-dependent phase caused by orbital evolution. Usually we deal with it by post-Newtanion(PN) approximation. Here we are not going to give too much explain. Because this term will be eliminated when we do the following math, so its concrete expression will not affect the final result. Here we just need to know that it is a function of the coalescence time $t_{c}$, the phase $\phi_{c}$ when emitted, $M_{z}$, $f$, $\eta$.

There are five unknown parameters, namely: $M_{z}$, $\eta$, $t_{c}$, $\phi_{c}$, $d_{L}$. $d_{L}$ is the only parameter that has nothing do with the own property of binary system. For the convenience of calculating, we just take account of equal mass NS binary system with $1.4M_{\bigodot}$ ,and set $t_{c}=0,\phi_{c}=0$. Then $M_{z}=1.22(1+z)M_{\bigodot},\eta=0.25$. The observation of GW can tell us no information about reshift, which means we should still resort to electromagnetic observation to find out corresponding redshift. Cutler and Holz has demonstrated its technological viability(\citealt{2009PhRvD..80j4009C}).

\begin{figure}[htb]
\center{\includegraphics[width=9cm]  {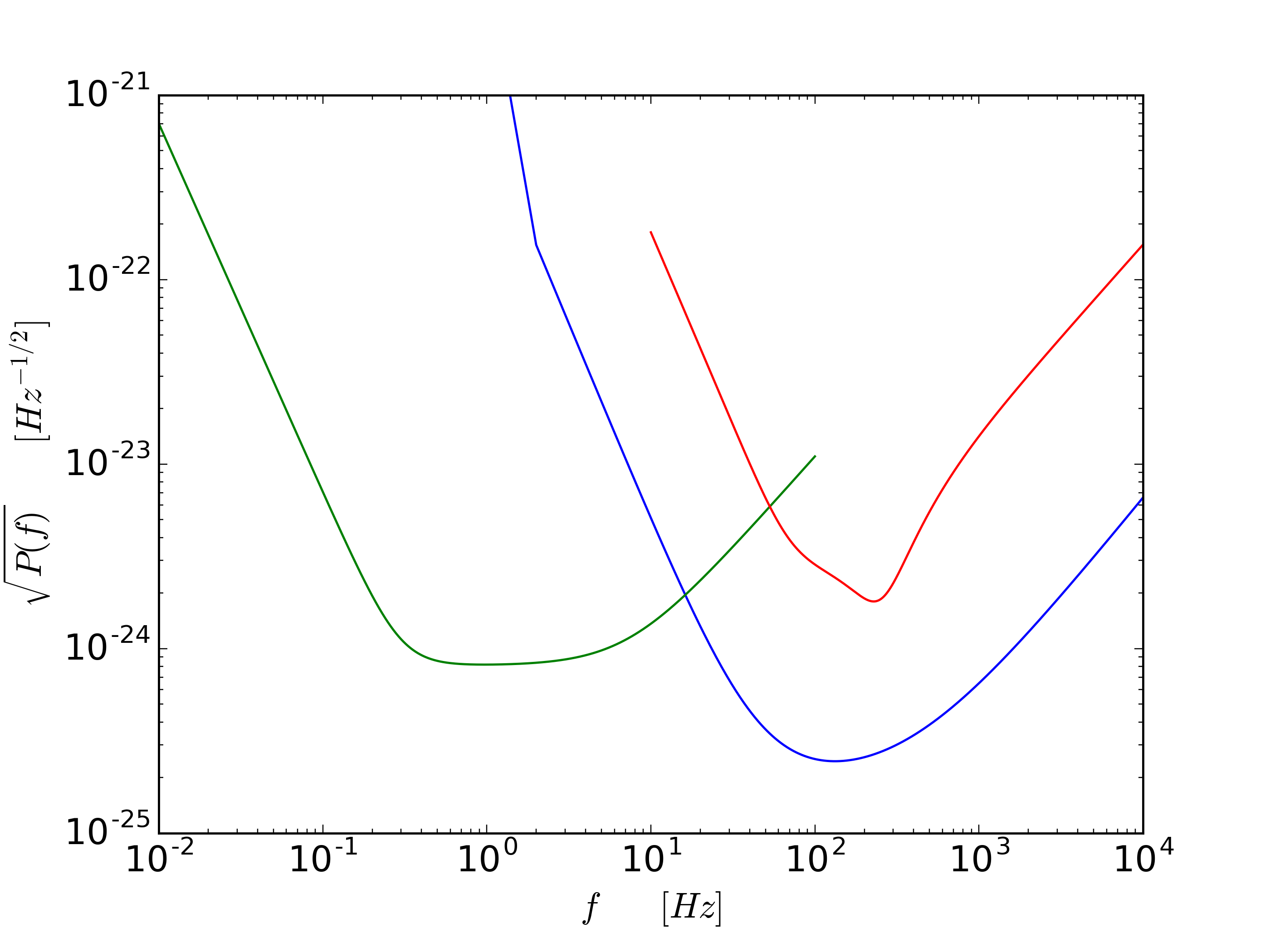}}
\caption{\label{2} The noise power spectrum. Green curve represents $P_{1}(f)$ for DECIGO, blue curve represents $P_{2}(f)$ for ET, red curve represents $P_{3}(f)$ for Adv-LIGO respectively}
\end{figure}

The estimate of error of $d_{L}$ is based on Fisher matrix. And the Fisher matrix here is given by
\[ \Gamma_{ab}=4 Re \int_{f_{min}}^{f_{max}}  \dfrac{ \partial_{a} \tilde{h}^{*}_{i}(f)   \partial_{b} \tilde{h}_{i}(f)    }{P(f)}  df  \]
$\partial_{a}$ means derivative with respect to parameter $\theta_{a}$. For DECIGO, who has eight interferometric signals, $\Gamma_{ab}$ should multiplied by 8. We have set values to parameters expect for $d_{L}$. So the only parameter in $\Gamma_{ab}$ is  $d_{L}$. P(\emph{f}) is the noise power spectrum. And the $P(f)$ for DECIGO, ET and Adv-LIGO is shown in Fig. \ref{2}. Here we give the expression of each detector's noise curve,$P_{1}(f),P_{2}(f),P_{3}(f)$ respective for DECIGO, ET and Adv-LIGO.\\

\textbf{DECIGO}: DECIGO is the acronym for Deci-Hertz Interferometer Gravitational Wave Observatory. DECIGO is a planed space-based GW observation aimed at $0.1~10Hz$ frequency region. Its configuration is still to be decided. Here we adopt the following parameters in its configuration: the arm length $1000km$, the output laser power $10W$ with wavelength $\lambda=532nm$, the mirror diameter $1m$ with its mass $100kg$, and the finesse of FP cavity 10. Its noise curve is(\citealt{2006CQGra..23S.125K})
\[
\begin{split}
P_{1}(f)=&6.53\times10^{-49}[1+(\dfrac{f}{7.36Hz})^{2}  \\
&+4.45\times10^{-51}\times (\dfrac{f}{1Hz})^{-4}\times \dfrac{1}{1+(\dfrac{f}{7.36Hz})^{2}}  \\
&+4.94\times10^{-52}\times (\dfrac{f}{1Hz})^{-4}] Hz^{-1}
\end{split}
\]

\textbf{ET}: ET is a third generation GW detector, whose design is not finished. Here we just consider the simplest case with 10 km arms.  We adopt the expression given by Keppel and Ajith's fitting used(\citealt{2010PhRvD..82l2001K})
\[
\begin{split}
P_{2}(f)= &10^{-50}[2.39\times10^{-27}(\dfrac{f}{100Hz})^{-15.64}   +0.349\times(\dfrac{f}{100Hz})^{-2.145}  \\ &+1.76\times(\dfrac{f}{100Hz})^{-0.12}+0.409\times(\dfrac{f}{100Hz})^{1.1}    ]^{2}   Hz^{-1}
\end{split} \]

\textbf{Adv-LIGO}: Adv-LIGO is an available second generation detector whose optimal sensitivity band match with the frequency window of GW from NS binary system . The first run of Adv-LIGO did not reach its design sensitivity. Here we use the noise curve fitted by [\citealt{2005PhRvD..71h4008A}]. It is not an accurate expression, but an approximation of the original curve given by [\citealt{2002gr.qc.....4090C}].
\[
\begin{split}
P_{3}(f)= &10^{-49}[ (\dfrac{f}{215Hz})^{-4.14}-5\times(\dfrac{f}{215Hz})^{-2}  \\
                    &+111\times( \dfrac{1-(\dfrac{f}{215Hz})^{2}+(\dfrac{f}{215Hz})^{4} /2 }{  1+(\dfrac{f}{215Hz})^{2}/2}   )     ]   Hz^{-1}
\end{split} \]

In the expression of $\Gamma_{ab}$, the lower cutoff of frequency, $f_{min}$, is a function of observation time $T_{obs}$.

\[ f_{min}=0.233( \dfrac{1M_{\bigodot}}{M_{z}}  )^{5/8}(   \dfrac{1yr}{T_{obs}} )^{3/8}Hz \]

\begin{figure}[htb]
\center{\includegraphics[width=9cm]  {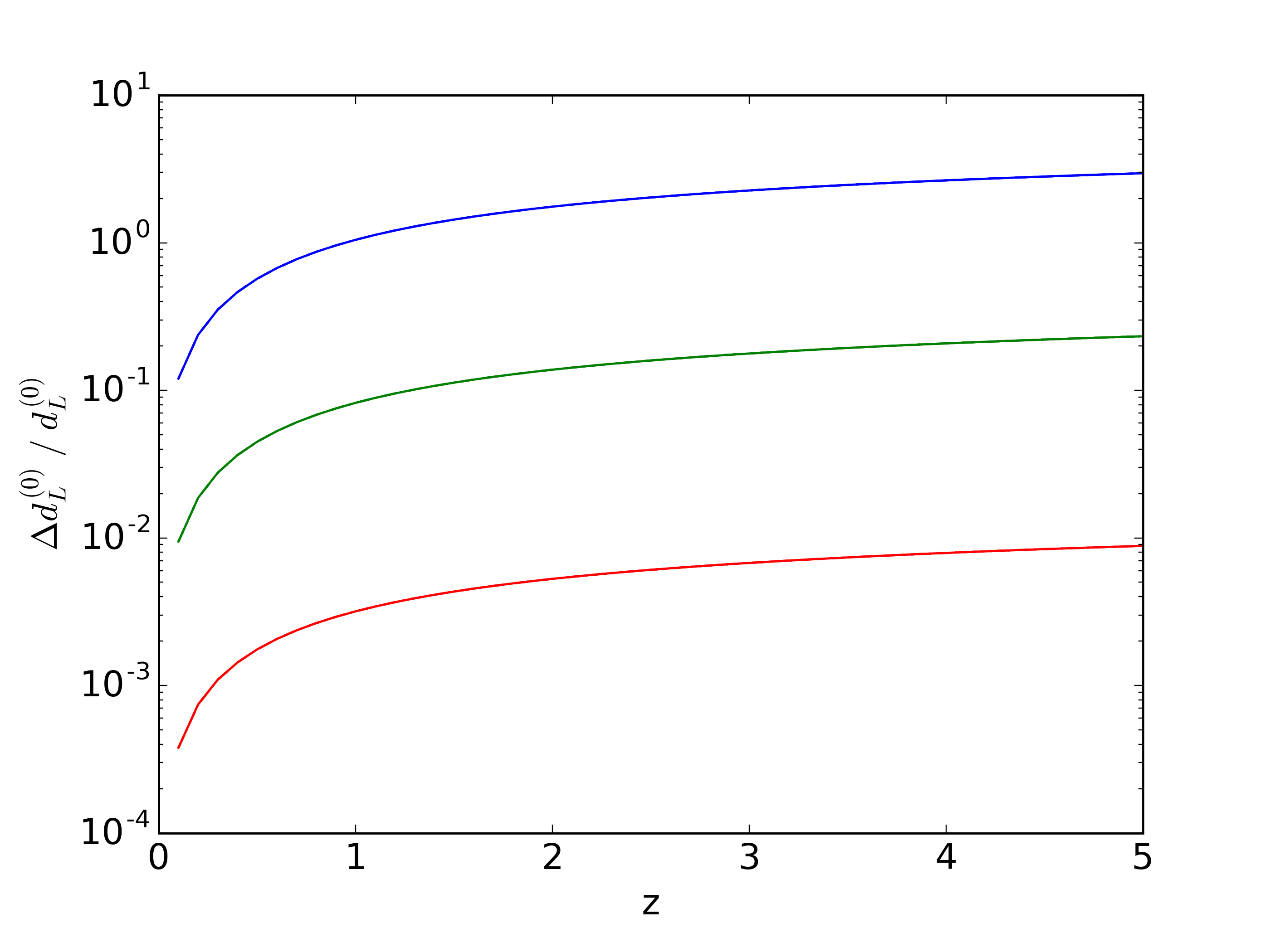}}
\caption{\label{3} The distance accuracy of $d_{L}$ by three devices. Different colors denote different devices, red for DECIGO, green for ET, and blue for Adv-LIGO.}
\end{figure}

\begin{figure}[htb]
\center{\includegraphics[width=9cm]  {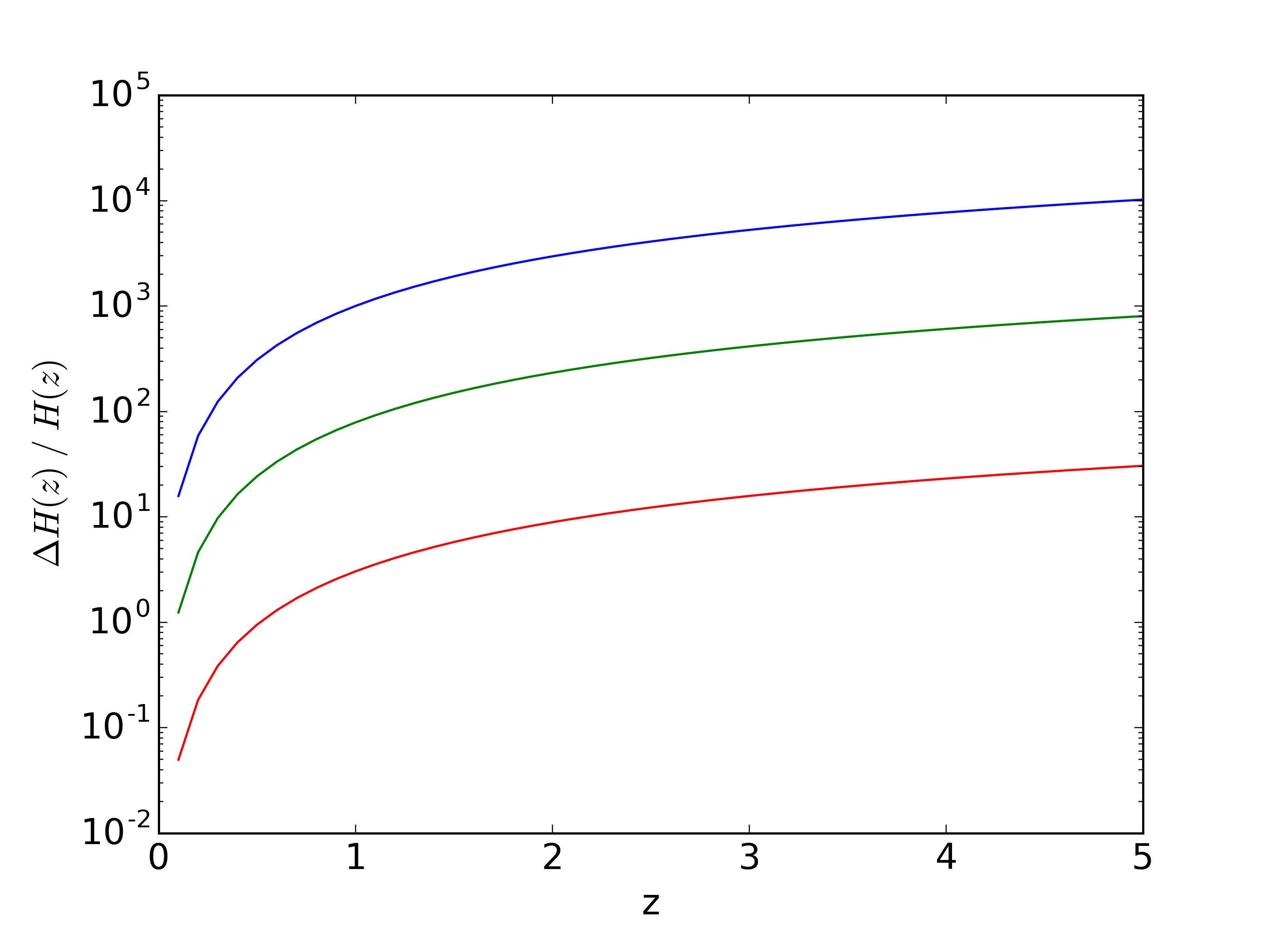}}
\caption{\label{4} The value of $\Delta H(z)/H(z)$ by only one source, red for DECIGO, green for ET, blue for Adv-LIGO.  }
\end{figure}

In the case of our paper, for a given $T_{obs}$, $f_{min}$ changes little with $M_{z}$, which is always in the high strain noise region. It makes no big difference to the result of the integral. A reasonable appointment of the value of $f_{min}$ will work. But for prudence, we just take the original expression of $f_{min}$ when calculate the integral. And the higher cutoff, $f_{max}$, can be decided by the property of the integrand. When the value of integrand goes comparatively small, its contribution to $\Gamma_{ab}$ can be neglected. Setting a upper limit of the integrating region is all right for the calculation. For the reason of their property of integrand, we set the $f_{max}$ of DECIGO, ET, Adv-LIGO respectively as $100Hz$, $10000Hz$ and $10000Hz$\\
The 1-sigma error of parameter is
\begin{center}
   $ \dfrac{\Delta d_{L}^{(0)}}{d_{L}^{(0)}} =\Delta\theta_{a}=\sqrt{\{\Gamma^{-1}\}_{aa}}$
\end{center}
For different observation time, ie: 1 year, 3 years, 10 years, the 1-sigma error estimate of $d_{L}$, which arising from instrumental noise, is showed in Fig. \ref{3}. we use $\sigma_{instr}$ to denote it. For a given device, no matter it is DECIGO, ET or Adv-LIGO, the accuracy of $d_{L}$ is all the same even for different observation time. It makes no difference for the error no matter how long the observation continues. It is mainly because that the error is due to the property of device, having nothing to do with observation time. Then we can calculate the $\Delta H(z)/H(z)$ by a simple multiply. The value of $\Delta H(z)/H(z)$ by analyzing the GW function of a NS binary system at a given redshift is shown is Fig. \ref{4}. As we can see, the accuracy is far from desirable. We need to take measure to narrow down the error.   \\

\subsection{H(z) error}

In last subsection, we already calculate the $H(z)$ relative error from a given NS binary system. The mean error will statistically abate if we have many independent sources. We can observe many NS binary system at the same redshift, which will lead us to a remarkably reduced error of $H(z)$. But to what degree can we reduce the error? First we need to figure out the number distribution of NS binary system, namely N, at different redshift. The distribution of NS binary system can be described and estimated. We are going to make use of it to estimate the accuracy of $H(z)$ in this subsection.
According to Ref[\citealt{2006PhRvD..73d2001C}], the fitting of NS-NS merger rate presents us the following mathematic describe:
\begin{center}
    $\dot{n}(z)=\dot{n}_{0}s(z)$, $s(z)=\begin{cases} 1+2z, &z\leq1 \\ 0.75(5-z), &1<z<5 \\ 0, &z\geq5   \end{cases}$

\end{center}
where the s(z) is estimated from star formation history inferred from UV luminosity, and the $\dot{n}_{0}$ represents the merger rate at present time. Then $\Delta N$, the number of NS-NS merger at redshift bin $\Delta z$, is expressed by:
    \[ \Delta N(z)=T_{obs}  \int_{z-\frac{\Delta z}{2}}^{z+\frac{\Delta z}{2}} { \dfrac{\dot{n}(z')}{1+z'}\times 4\pi[\dfrac{d_{L}(z')}{1+z'}]^{2}\times \dfrac{1}{H(z')}    dz'} \]

\begin{figure}[htb]
\center{\includegraphics[width=9cm]  {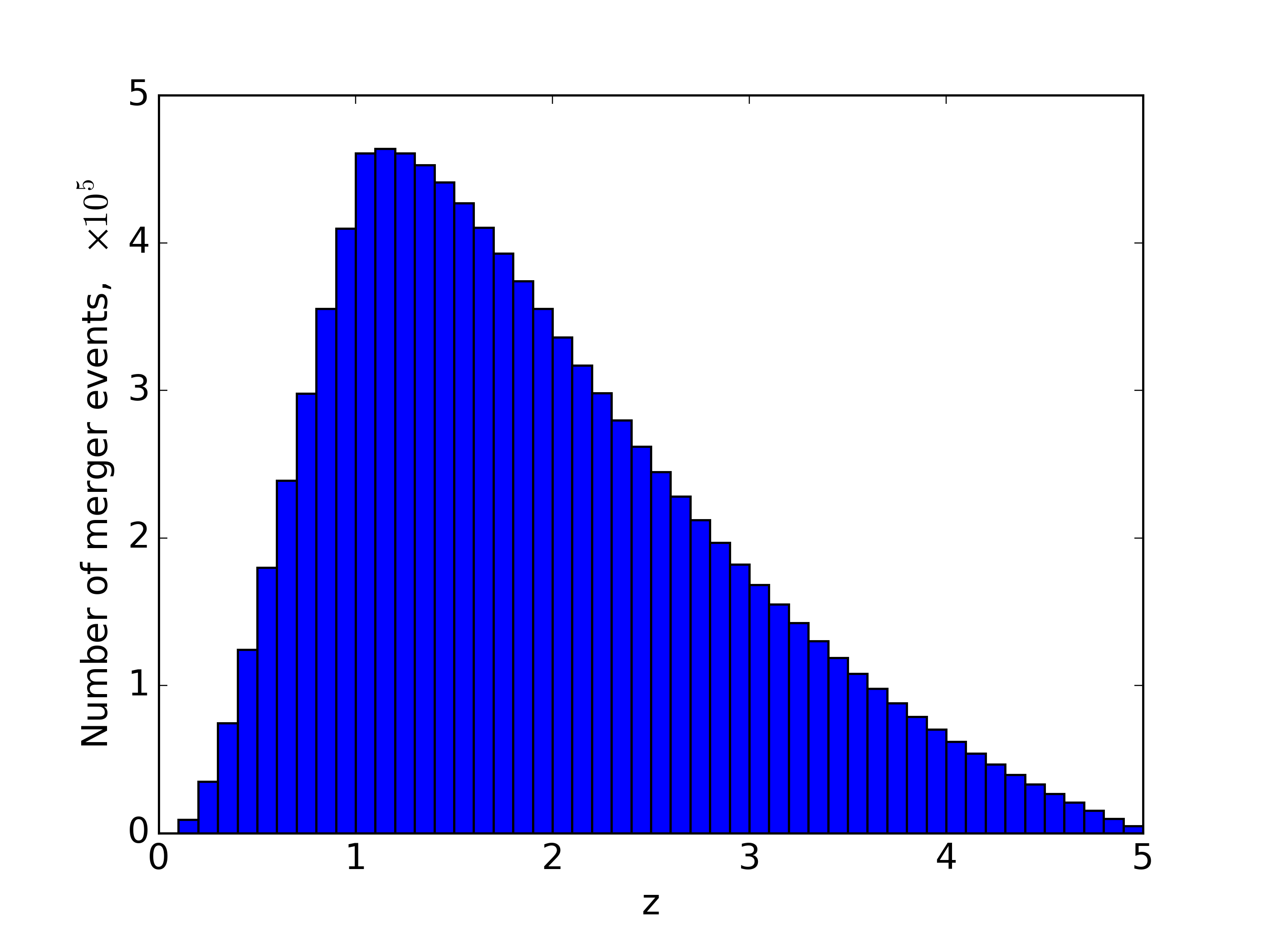}}
\caption{\label{5} The number of merger events at every 0.1 z redshift, for a 10-year observation}
\end{figure}

Current study doesn't provide solid evidence of the exact value of $s(z)$ and $\dot{n}_{0}$. But we can take a reasonable estimate . After all, we are aimed at evaluating the method, not launching an actual observation here. It is rational to set that $\dot{n}_{0}$ equals to the most recent estimate, $10^{-6}Mpc^{-3}yr^{-1}$, and $\Delta z$ equals to 0.1. Thus we get the estimation of 10-year observed number of NS binary system merger at different redshift, which is shown in Fig. \ref{5}.

\begin{figure}[htb]
\center{\includegraphics[width=9cm]  {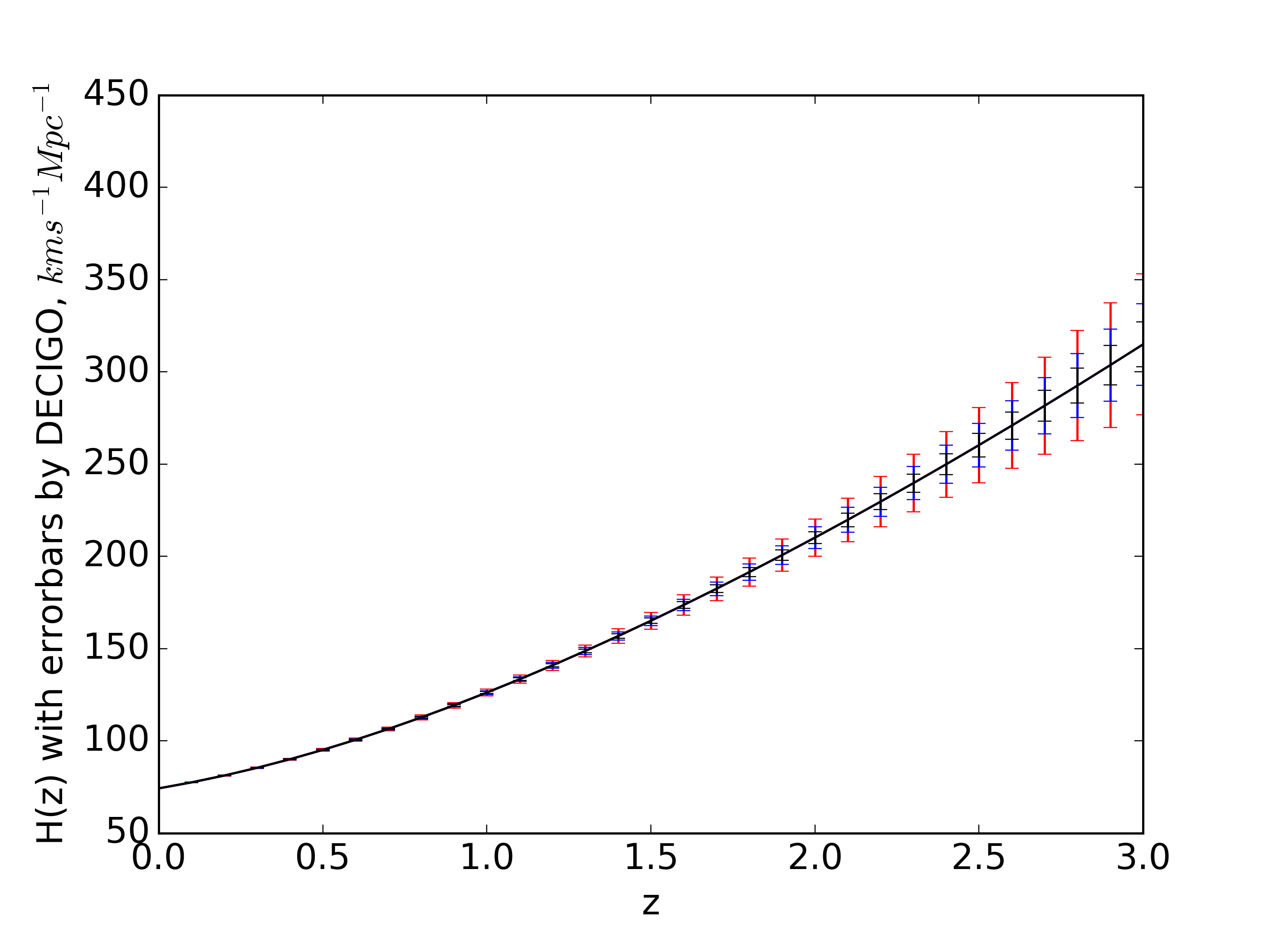}}
\caption{\label{6}  $H(z)$ with error bars by DECIGO under $\Lambda CDM$. Error bars of different colors represents different observation time, red for 1-year, blue for 3-year, black for 10-year observation respectively}
\end{figure}

\begin{figure}[htb]
\center{\includegraphics[width=9cm]  {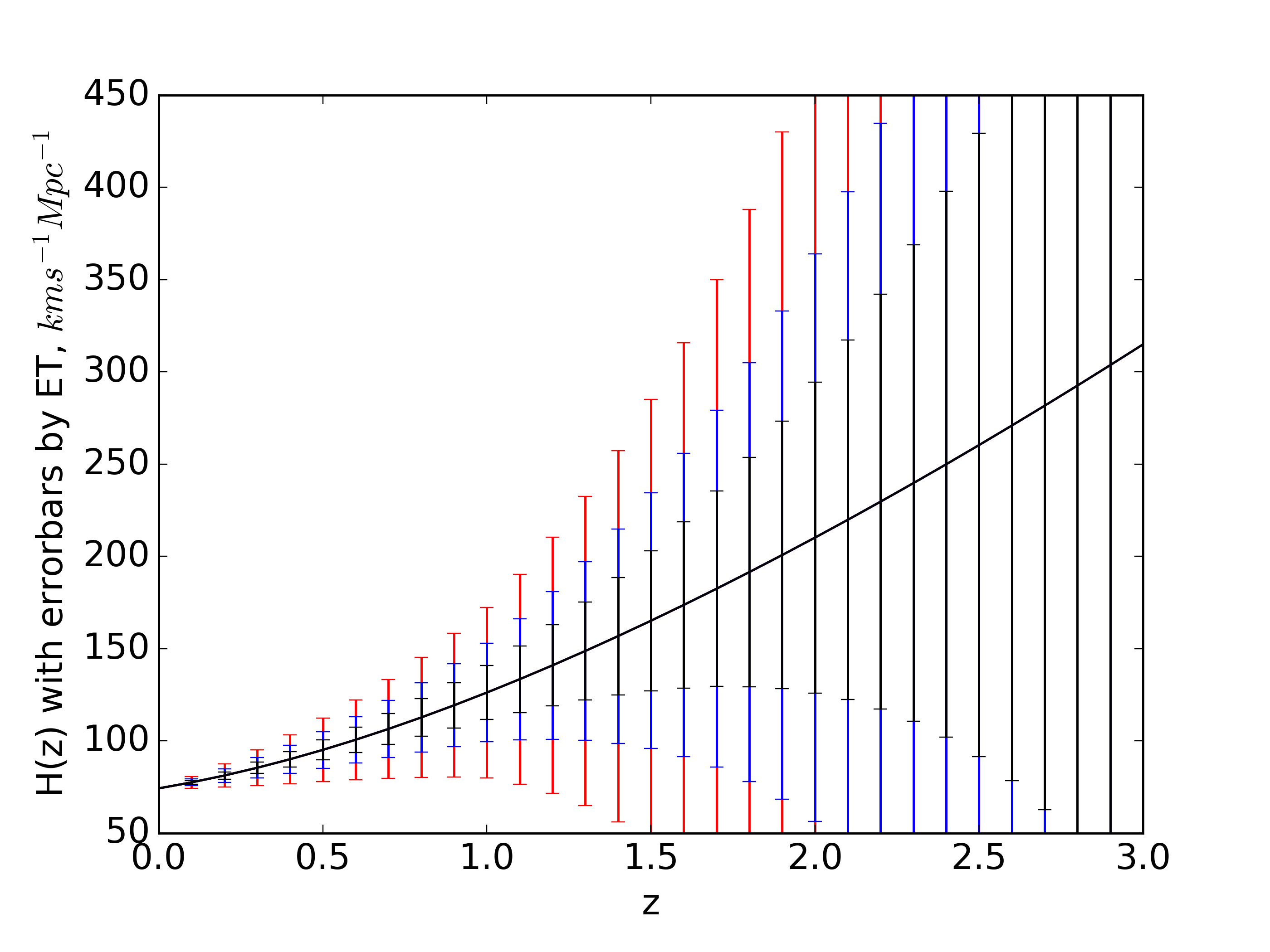}}
\caption{\label{7} Same as Fig. \ref{6}, but for ET}
\end{figure}

\begin{figure}[htb]
\center{\includegraphics[width=9cm]  {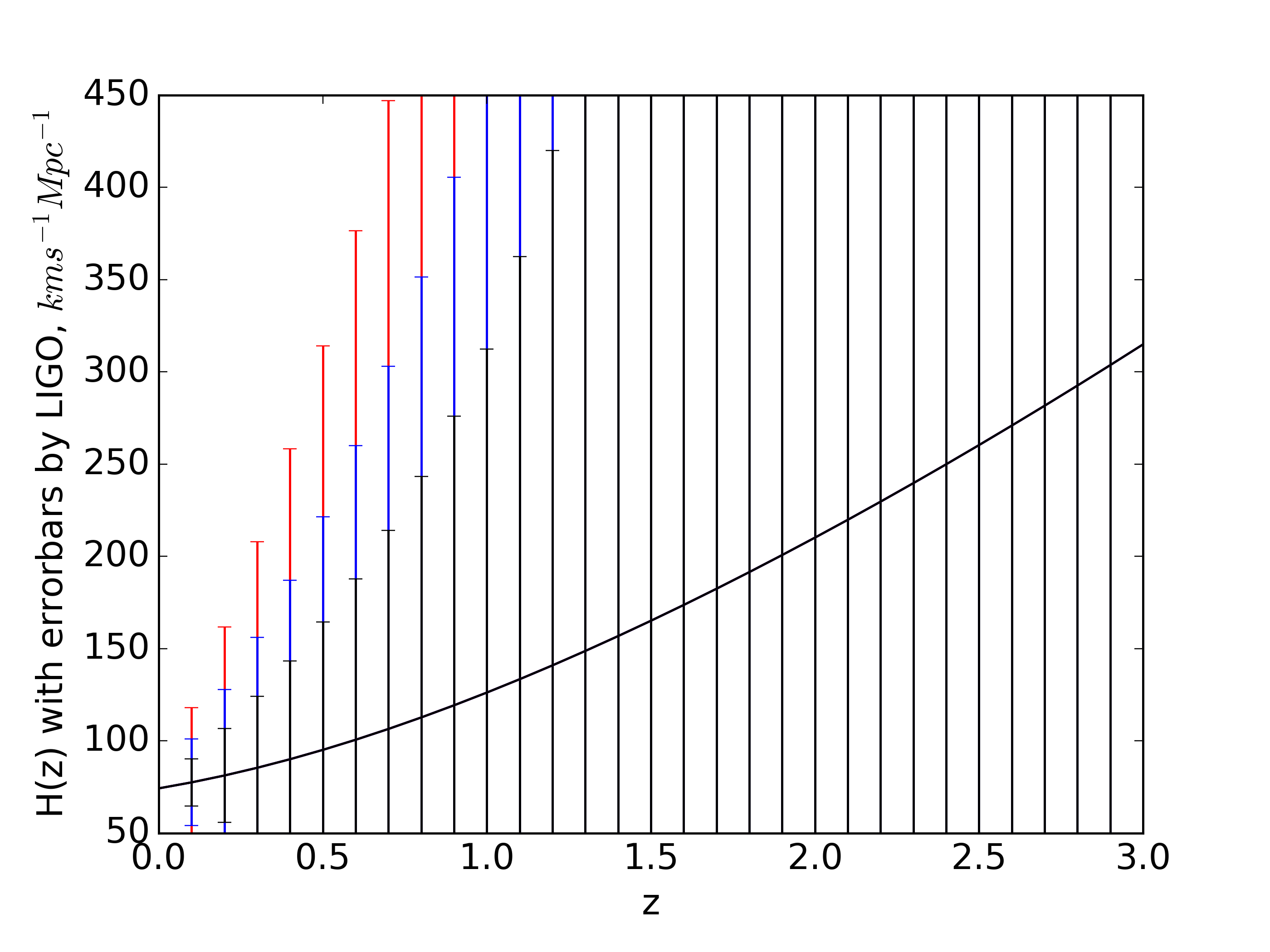}}
\caption{\label{8} Same as Fig. \ref{6}, but for Adv-LIGO}
\end{figure}

Given the fact that the total number of SN is just of hundred-magnitude by now, the observed number of NS-NS merger event is much larger than that of SN, showing a tremendous potential in reducing the mean error of $H(z)$. And from above equation, the number of NS-NS merger at fixed redshift increases with $T_{obs}^{1/2}$. The elongation of observation time can remedy the drawback of the device sensibility. This is also an advantage of the replacement of SN by NS-NS merger event. Combining with the information we get in last subsection, we are able to calculate the $H(z)$ error now. To make it intuitive, we convert relative error to absolute error of $H(z)$ under $\Lambda CDM$. We choose $\Lambda CDM$ as fiducial model is out of the consideration that $\Lambda CDM$ is mostly consistent with cosmological observation. The relative error of $H(z)$ by DECIGO ,ET and Adv-LIGO is shown in Fig. \ref{6}, Fig. \ref{7}, and Fig. \ref{8}, each for 1-year, 3-year, 10-year observation. The error by Adv-LIGO is a totally disaster, which basically has little application value in constraining cosmological parameters. The error by ET is a little better, especially at low redshift region, because ET is more sensitive than Adv-LIGO. DECIGO plays quite well in this method. For 10-year observation, the relative error is less than $1\%$ at the range $z\leq0.5$. When redshift reach 3, Due to the decreasing of the number of observed NS-NS merger event with redshift, the relative error of $H(z)$ was magnified, but still quite small. And the elongation of $T_{obs}^{1/2}$ shows a great ability in narrowing down the error.

\section{evaluation}

\subsection{simulate data}
\label{sec:morph}

The new method for detecting $H(z)$ has been proposed and a little error analysis has been done. The problem is how accurate could $H(z)$ observed by this way constrain cosmological parameters? Now it is just a method, which could not operate so far. We get no actual OHD by this way. But it doesn't necessarily mean we can do nothing about it. A reasonable and rational simulation would help our forecasting and evaluating a lot.\\
Since the $H(z)$ data by Adv-LIGO have a bad accuracy, we are just going to carry on no simulation and forecast for $H(z)$ data by Adv-LIGO here. ET can do some simulation and forecast. The problem is that the effect is a little bad, even worse than 38 OHDs. We do not plan to show it here, too. In following sections, DECIGO is the only one we discuss. we are going to follow the way Yuan Shuo once took to generate mock data(\citealt{2015JCAP...02..025Y}),
\begin{center}
    $H_{sim}=H_{\Lambda CDM}+H_{drift}$
\end{center}

\begin{figure}[htb]
\center{\includegraphics[width=9cm]  {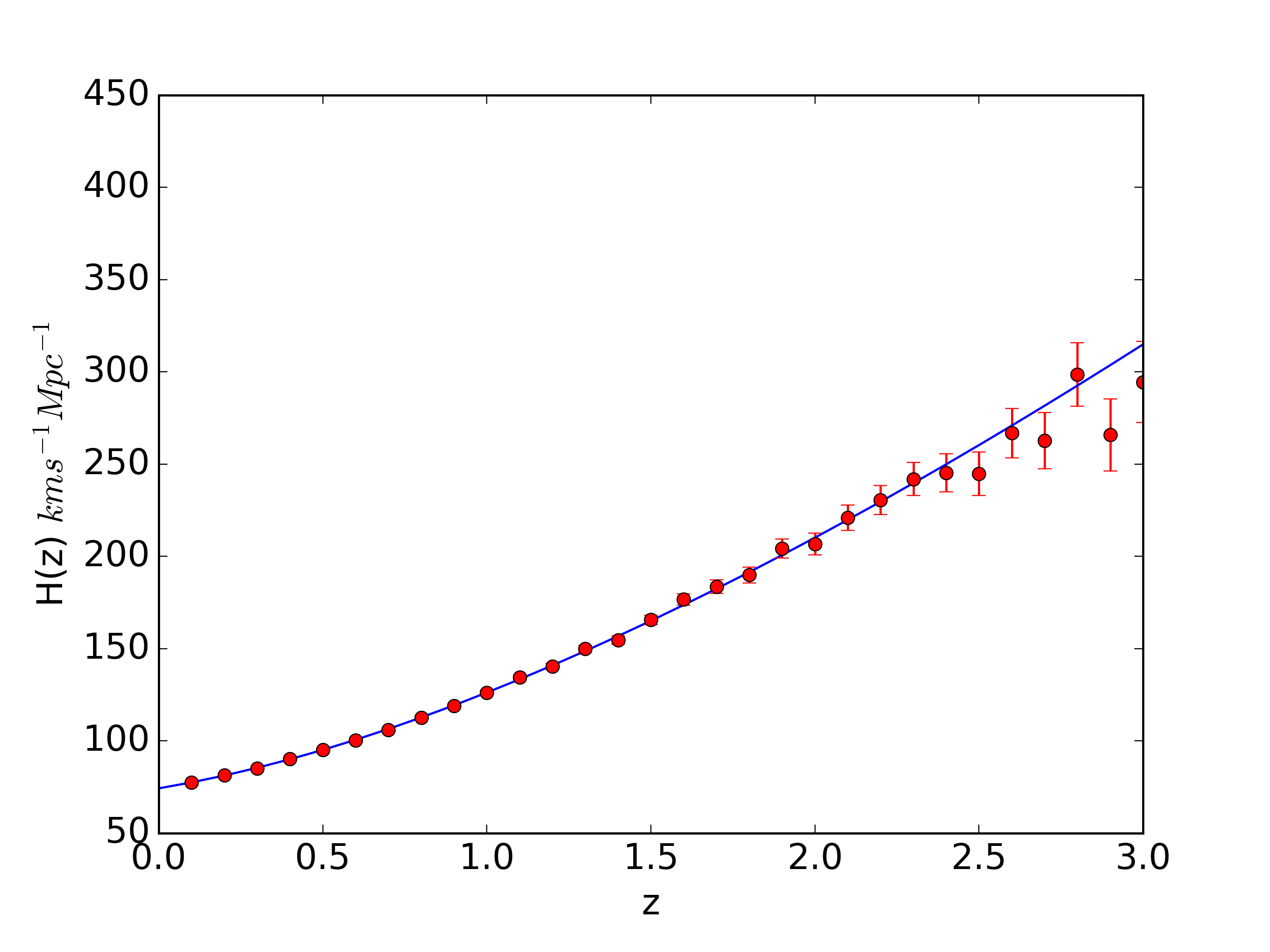}}
\caption{\label{9} Mock data. The blue curve denotes H(z) value under $\Lambda CDM$, while the dots with error bars represent simulation data }
\end{figure}

\begin{figure}[htb]
\center{\includegraphics[width=9cm]  {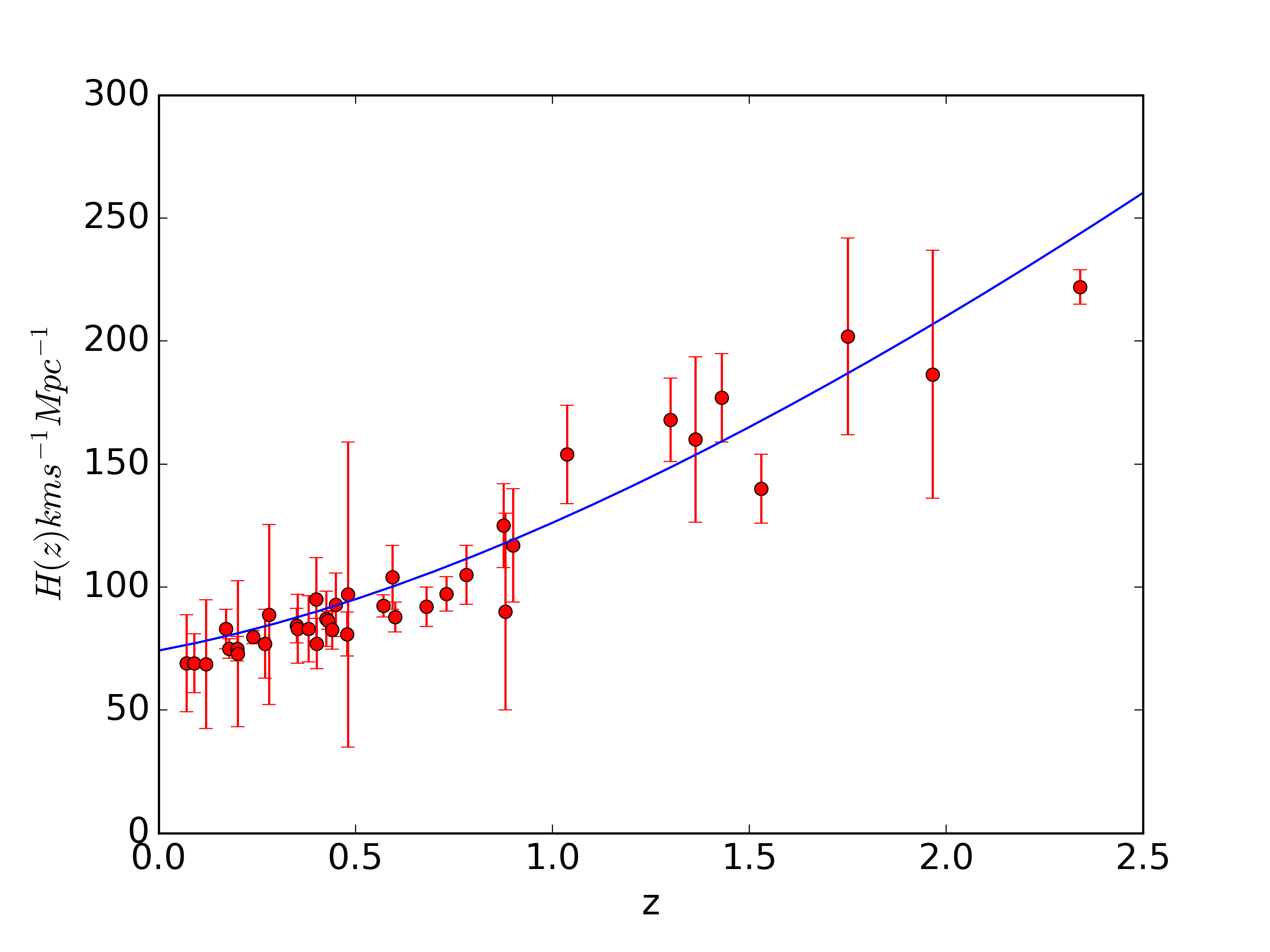}}
\caption{\label{10} 38 OHDs. The dots with error bars represent 38 available OHDs so far. For the purpose of illustrating, we also plot $H(z)$ value under $\Lambda CDM$, the blue curve   }
\end{figure}

We treat $H_{sim}$ as a drift, $H_{drift}$, based on the theoretical $H(z)$ value under $\Lambda$CDM, $H_{\Lambda CDM}$, caused by various errors. $H_{drift}$ is a random value under gaussian distribution, $N(0,\Delta H)$. $\Delta H$ is the calculated by relative error we get in last section. Using a piece of python code, we generate our mock $H_{sim}$ data of 3-year observation at very 0.1 z, shown in Fig. \ref{9}. We have got 38 OHDs up to now. The data were obtained by different ways from different groups(\citealt{2003ApJ...593..622J};\citealt{2005PhRvD..71l3001S};\citealt{2010JCAP...02..008S};\citealt{2012JCAP...07..053M};
\citealt{2016JCAP...05..014M};\citealt{2014RAA....14.1221Z};\citealt{2015MNRAS.450L..16M};\citealt{2009MNRAS.399.1663G};\citealt{2012MNRAS.425..405B};
\citealt{2013MNRAS.429.1514S};\citealt{2013MNRAS.431.2834X};\citealt{2015arXiv150702517M}). And their value has a distribution.  Fig. \ref{10} shows the $H_{\Lambda CDM}$ at every redshift and the 38 OHDs so far. As we can see, the OHD value goes up and down around the $H_{\Lambda CDM}$ at the same redshift, which justifies the validity of our simulation. \\

\subsection{forecasting of mock data}
\label{sec:velocity}

Now that we have got the 3-year-observation mock data, we can use them to constrain cosmological parameters. Before that, we need a criteria to evaluate the constraining ability of the dataset-Figure of Merit(FoM). We can define FoM in different ways, as long as its value can reflect how tightly or loosely the data constrain parameters. Here for the convenience of our analysis, we adopt the definition in Ref(\citealt{2006astro.ph..9591A}), the reciprocal of the area enclosed by the contour, coinciding with a specially appointed confidence region under gaussian distribution.\\

We choose the $\Lambda CDM $ as our prior model. In a standard $\Lambda CDM $ universe with a curvature term $\Omega_{k}=1-\Omega_{m}-\Omega_{\Lambda}$,the Hubble parameter is given by
\begin{center}
   $H(z)=H_{0}E(z);  E(z)=\sqrt{\Omega_{m}(1+z)^{3}+\Omega_{\Lambda}+\Omega_{k}(1+z)^{2}} $
\end{center}
The determination of $H_{0}$ has been carried on in different projects. Its value varies slightly from one project to another, but always around $70km s^{-1}Mpc^{-1}$. Freedman suggested $H_{0}=72\pm8 km s^{-1} Mpc^{-1}$. For 7-year WMAP observation, $H_{0}=73\pm3 km s^{-1} Mpc^{-1}$(\citealt{2007ApJS..170..377S}). In this paper, we take the most recent value $H_{0}=74.2\pm3.6 km s^{-1}Mpc^{-1}$(\citealt{2009ApJ...699..539R}). And the best value of $\Omega_{m}$,$\Omega_{\Lambda}$ we adopt is 0.27, 0.73 respectively, due to the coherence that they are consistent with the observations and the fact that we use these value to generate our simulation data. All the three parameters are assumed under gaussian distribution. By Bayes' theorem, the posterior probability density function of parameters given the data set \{$H_{i}$\} is :

\[ P( \Omega_{m},\Omega_{\Lambda}| \{H_{i}\} )=    \int P( \Omega_{m},\Omega_{\Lambda},H_{0} | \{H_{i}\} )  d H_{0} \]
\[   =  \int \ell( \{H_{i}\}  | \Omega_{m},\Omega_{\Lambda},H_{0} )  P(H_{0})  d H_{0} \]

where $\ell$ is the likelihood and $P(H_{0})$ is the prior probability density function of $H_{0}$. And the expression of $\ell$ is given by
\[ \ell( \{H_{i}\} | \Omega_{m},\Omega_{\Lambda},H_{0} )=\left( \prod_{i} \dfrac{1}{\sqrt{2\pi\sigma_{i}^{2}}} \right) exp \left(-\dfrac{\chi^{2}}{2}\right) \];
\[ \chi^{2}=\sum_{i}  \dfrac{[ H_{0}E(z) -H_{i}]^{2} }{\sigma_{i}^{2}}   \]
and $\sigma_{i}$ is the uncertainty of the data $H_{i}$. Then the integral can be worked out for a given $P(H_{0})$.
There would be a point in the parameter space maximizing the probability density, $P_{max}$. Because of what we have described in last paragraph, such a point in this forecasting would be \{0.27, 0.73, 74.2\}. The formula

\begin{center}
    $P=P_{max} exp \left(-\dfrac{\Delta \chi^{2}}{2}\right) $
\end{center}

means the contour of a given confidence region, which corresponds to the value of $\Delta \chi^{2}$. We have three parameters, $\Omega_{m},\Omega_{\Lambda},H_{0}$, here. $\Delta \chi^{2}$ is statistically set to 2.3, 6.17, 11.8 respectively for $1\sigma, 2\sigma, 3\sigma$ confidence region. For a direct comparing and understanding, here we choose $2\sigma$ confidence region, namely $\Delta \chi^{2}=6.17$, when calculate FoM.

\begin{figure}[htb]
\center{\includegraphics[width=9cm]  {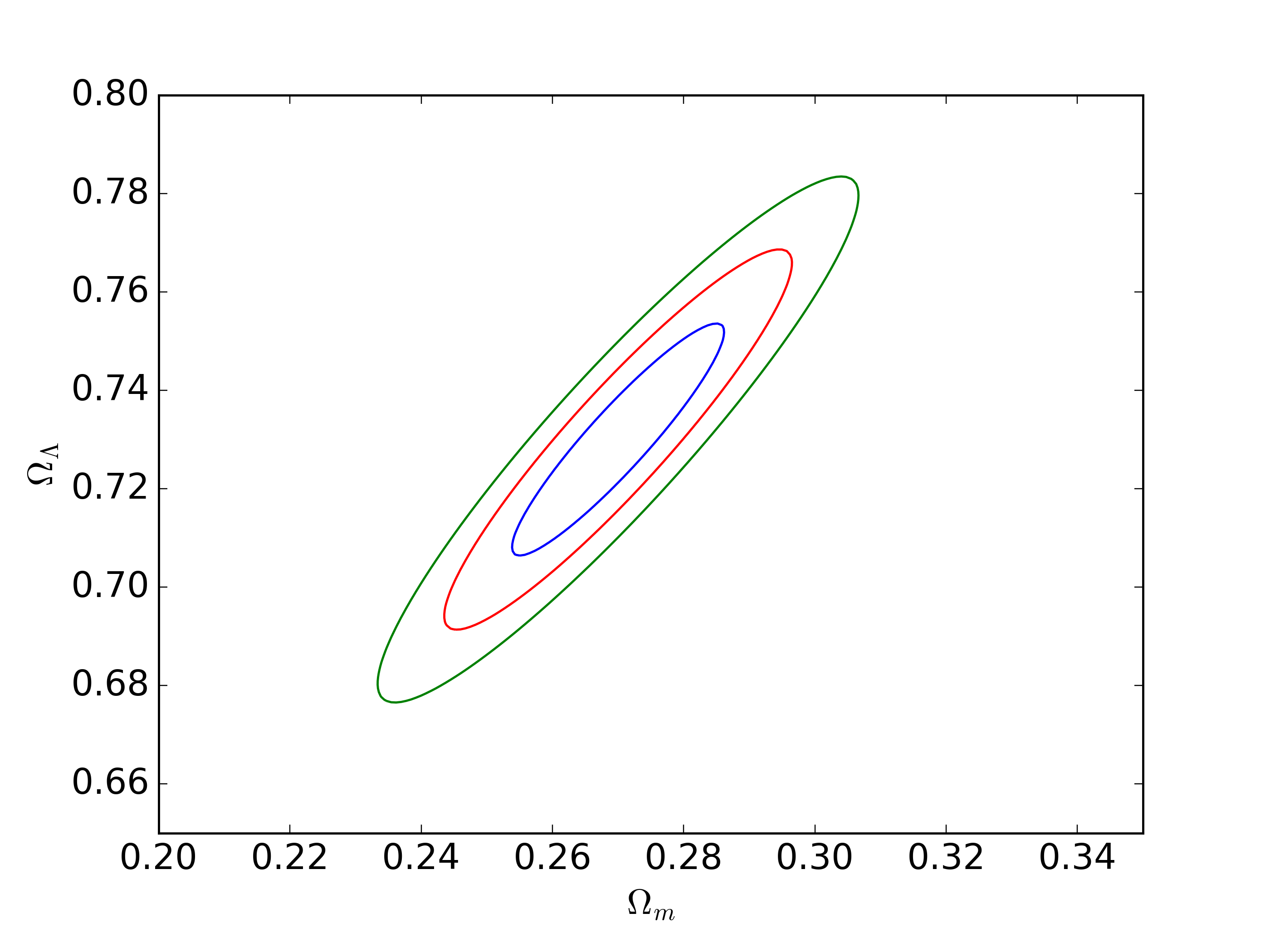}}
\caption{\label{11} Constraint on $\Omega_{m}$ and $\Omega_{\Lambda}$ by 3-year-observation. The blue, red, green curve denote $1\sigma, 2\sigma, 3\sigma$ confidence region, The FoM of simulation data is 834.9}
\end{figure}

To estimate the FoM, we take the Fisher Matrix forecast technique in [\citealt{2003moco.book.....D}], and

\begin{figure}[htb]
\center{\includegraphics[width=9cm]  {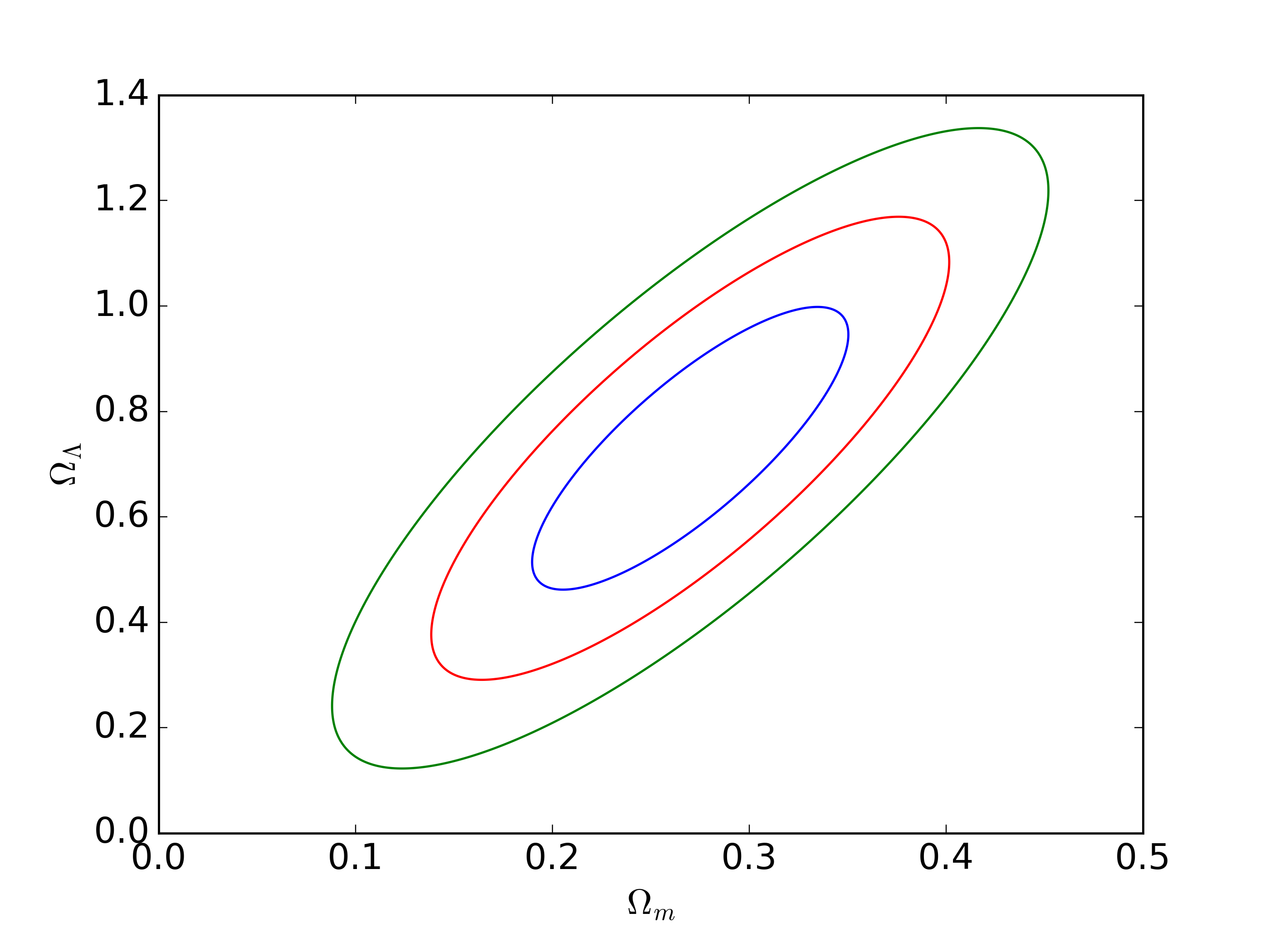}}
\caption{\label{12} Same as Fig. \ref{11}, but for 38 actual OHDs. The FoM here is 9.3}
\end{figure}

\begin{figure}[htb]
\center{\includegraphics[width=9cm]  {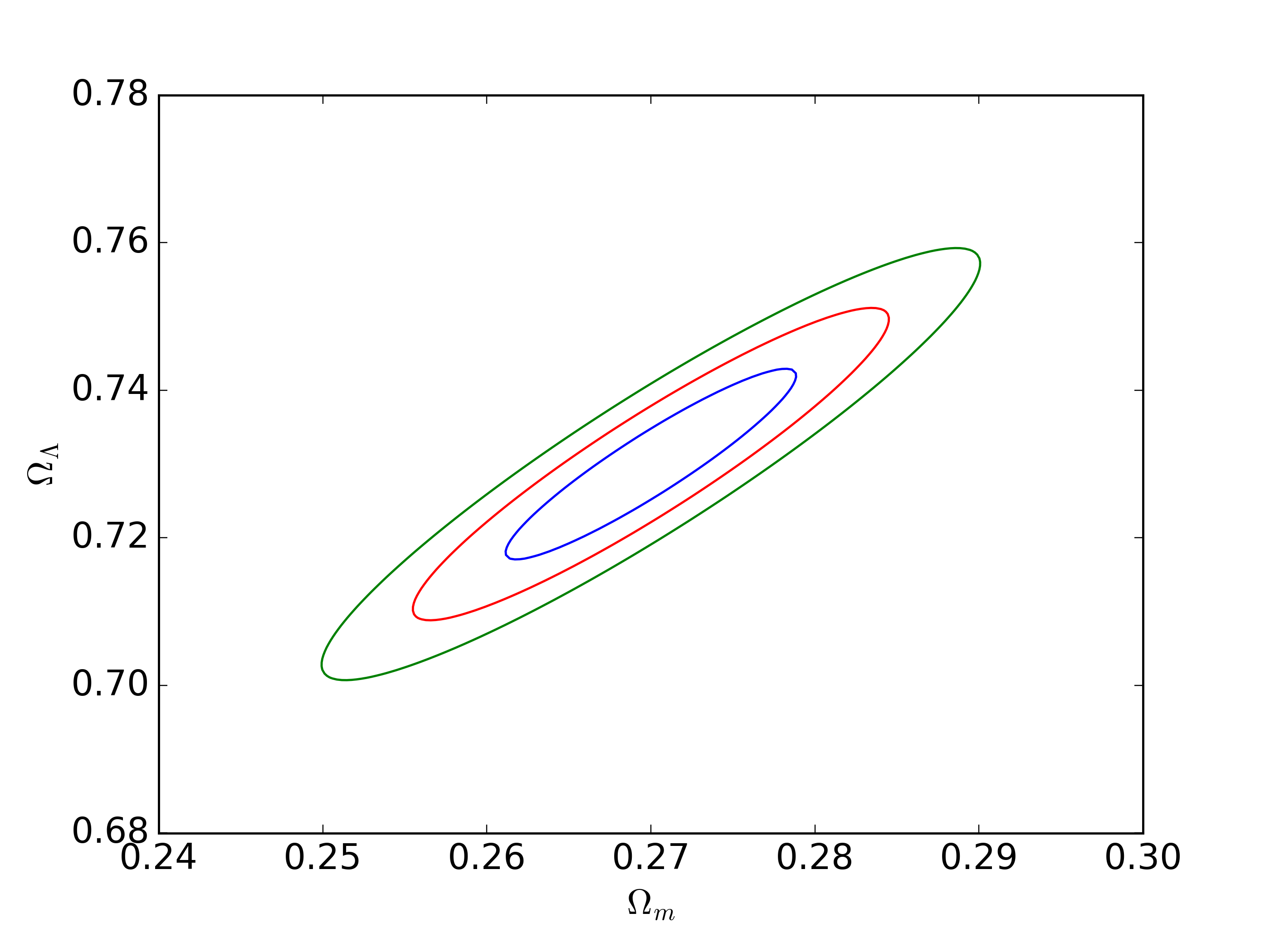}}
\caption{\label{13} Same as Figure 10, but for 10-year-observation simulation. The FoM here is 2783.1}
\end{figure}

\begin{center}
   $F_{ij}=\dfrac{1}{2}\dfrac{\partial^{2}\chi^{2}}{\partial\theta_{i}\partial\theta_{j}}$
\end{center}
the value of matrix elements is taken at the most-likely value of parameters. It is easy to calculate the Fisher matrix, a $3\times3 $ matrix in parameter space \{ $\Omega_{m},\Omega_{\Lambda},H_{0}$\}. But we need to marginalize it to obtain the FoM in subspace \{$\Omega_{m},\Omega_{\Lambda}$\}. The marginalization is not intricate: inverse the matrix and remove the row and column to be marginalized, then inverse the reduced matrix again. Let's denote the marginalized Fisher matrix by $\tilde{F}$. Then the contour in subspace can be given by
\begin{center}
   $(\Delta\theta)^{T} \tilde{F} \Delta\theta=\Delta\chi^{2}; \Delta\theta=\theta-\theta_{best-value}$
\end{center}
$\Delta\theta$ is the deviation from the beat value of the parameters. The contour is shown in Fig. \ref{11}. As we can see, the contour is an ellipse, which is consistent with the equation of $\tilde{F}$. For a more direct and concrete comparison, we cite the 38 OHDs here. Their constraint on $\Omega_{m}$ and $\Omega_{\Lambda}$ is shown in Fig. \ref{12}. Apparently, the constraint of the mock data on parameters is much tighter, compared with that of available OHDs, which has an significant improvement on "accurate cosmology". The simulation and forecasting of 10-year-observation just carries out in the same way. We are going to skip the  elaborate explanation here. As Fig. \ref{13} shows, its  constraint on cosmological parameters is even much tighter, implying a consequent higher FoM value.\\
When calculating FoM, we take $\Delta\chi^{2}$ as 6.17. The enclosed area is $\pi/ \sqrt{det(\tilde{F}/\Delta\chi^{2})}$. So FoM, the reciprocal of the area, is
\begin{center}
   $FoM=\dfrac{ \sqrt{det(\tilde{F}/\Delta\chi^{2})}}{\pi}$
\end{center}
By this way, we have the FoM value of mock data, which is about 834.9, while the FoM of 38 OHDs is just about 9.3. It is a remarkable improvement. For 10-year-observation mock data, the FoM has a farther improvement, reaching 2783.1. We have enough reason to look forward to the excellent application of $H(z)$ data by this method.

\section{Discussion and Conclusions}

In this paper, we mainly evaluate the quality of $H(z)$ data by GW standard siren method of several GW detection plans, whose optimal frequency locate around the frequency window of GW from typical BN system. We calculate the relative error of $H(z)$ on three devices, DECIGO and ET and Adv-LIGO. Though the sensitivity of the three devices is almost of the same order of magnitude, the $H(z)$ error of DECIGO is quite optimistic while that of other two is far from satisfying. It is because of the term in the integrand of $\Gamma$, $f^{-7/3}/P(f)$. The most sensitive frequency of Adv-LIGO and ET is of ~kHz, while DECIGO sensitive on ~10Hz, which is the exact reason leading to a difference on the integral. In low frequency region, the $P(f)$ of DECIGO is smaller, its contribution to integral is bigger. When it goes to high frequency region. Though $P(f)$ of ET and Adv-LIGO is small, $f^{-7/3}$ diminishes the value contributed by high frequency region to $\Gamma$. So the integral of ET and Adv-LIGO is smaller, leading a comparatively bigger error. But it does not mean that $H(z)$ data by this method is a dead end or of no meaning, which is justified by the forecasting of DECIGO-based $H(z)$ data. If the sensitivity of Adv-LIGO or ET is sightly improved, or just move the most sensitive frequency to a lower region, the error of $H(z)$  will be comparable that by DECIGO.

Considering the absence of real $H(z)$ data by DECIGO, we simulate $H(z)$ and the data show an alluring  constraining ability on cosmological parameters. After all, we are aimed at evaluating the viability and quality of $H(z)$ data by GW standard siren method, not putting the method into actual operation. We find that the FoM of mock data shows a huge improvement when compared with that of 38 actual OHDs. For contrast, the FoM is 9.3 for 38 OHDs, 834.9 for 3-year-observation, 2783.1 for 10-year-observation. The tight constraint of mock data and the FoM of the contour indicates a bright further of detecting $H(z)$ data by this method. It will bring "accurate cosmology" to next stage.

It is well worth to point out that at current stage, the range of Adv-LIGO's detecting ability for NS binary system is just 70 or 80Mpc. This range is much smaller than what we assumed here. In the further, if we want to detect $H(z)$ by this way, a farther detecting range is necessary, implying a lower strain noise. A lower strain noise will eventually lead to a more accurate $H(z)$ data. So if one day we can put GW standard sirens into practice, the $H(z)$ data we get should be more accurate than we simulate in this paper, and the constraining on cosmological parameters will be tighter.

\section*{Acknowledgements}

We thank to the many helpful suggestion from teachers and classmates. Especially Duan Xiaowei and Yu Hai help a lot. The discuss with them improve this work.
This work was supported by the National Science Foundation of China (Grants No. 11573006, 11528306), the Ministry of Science and Technology National Basic Science program (project 973) under grant No. 2012CB821804. And Y. K. Tang acknowledges financial support from the National Natural Science Foundation of China (Grant Nos. 11303003 and 11673005) and from the Shandong Provincial Natural Science Foundation of China (Grant No. ZR2013AM002)

\bibliographystyle{aasjournal}
\bibliography{APJ-paper}

\begin{thebibliography}{}
\expandafter\ifx\csname natexlab\endcsname\relax\def\natexlab#1{#1}\fi

\bibitem[{{Abbott} {et~al.}(2016){Abbott}, {Abbott}, {Abbott}, {Abernathy},
  {Acernese}, {Ackley}, {Adams}, {Adams}, {Addesso}, {Adhikari}, \&
  et~al.}]{2016PhRvL.116f1102A}
{Abbott}, B.~P., {Abbott}, R., {Abbott}, T.~D., {et~al.} 2016, Physical Review
  Letters, 116, 061102

\bibitem[{{Albrecht} {et~al.}(2006){Albrecht}, {Bernstein}, {Cahn}, {Freedman},
  {Hewitt}, {Hu}, {Huth}, {Kamionkowski}, {Kolb}, {Knox}, {Mather}, {Staggs},
  \& {Suntzeff}}]{2006astro.ph..9591A}
{Albrecht}, A., {Bernstein}, G., {Cahn}, R., {et~al.} 2006, ArXiv Astrophysics
  e-prints, astro-ph/0609591

\bibitem[{{Arun} {et~al.}(2005){Arun}, {Iyer}, {Sathyaprakash}, \&
  {Sundararajan}}]{2005PhRvD..71h4008A}
{Arun}, K.~G., {Iyer}, B.~R., {Sathyaprakash}, B.~S., \& {Sundararajan}, P.~A.
  2005, \prd, 71, 084008

\bibitem[{{Blake} {et~al.}(2012){Blake}, {Brough}, {Colless}, {Contreras},
  {Couch}, {Croom}, {Croton}, {Davis}, {Drinkwater}, {Forster}, {Gilbank},
  {Gladders}, {Glazebrook}, {Jelliffe}, {Jurek}, {Li}, {Madore}, {Martin},
  {Pimbblet}, {Poole}, {Pracy}, {Sharp}, {Wisnioski}, {Woods}, {Wyder}, \&
  {Yee}}]{2012MNRAS.425..405B}
{Blake}, C., {Brough}, S., {Colless}, M., {et~al.} 2012, \mnras, 425, 405

\bibitem[{{Bonvin} {et~al.}(2006{\natexlab{a}}){Bonvin}, {Durrer}, \&
  {Gasparini}}]{2006PhRvD..73b3523B}
{Bonvin}, C., {Durrer}, R., \& {Gasparini}, M.~A. 2006{\natexlab{a}}, \prd, 73,
  023523

\bibitem[{{Bonvin} {et~al.}(2006{\natexlab{b}}){Bonvin}, {Durrer}, \&
  {Kunz}}]{2006PhRvL..96s1302B}
{Bonvin}, C., {Durrer}, R., \& {Kunz}, M. 2006{\natexlab{b}}, Physical Review
  Letters, 96, 191302

\bibitem[{{Cutler} \& {Harms}(2006)}]{2006PhRvD..73d2001C}
{Cutler}, C., \& {Harms}, J. 2006, \prd, 73, 042001

\bibitem[{{Cutler} \& {Holz}(2009)}]{2009PhRvD..80j4009C}
{Cutler}, C., \& {Holz}, D.~E. 2009, \prd, 80, 104009

\bibitem[{{Cutler} \& {Thorne}(2002)}]{2002gr.qc.....4090C}
{Cutler}, C., \& {Thorne}, K.~S. 2002, ArXiv General Relativity and Quantum
  Cosmology e-prints, gr-qc/0204090

\bibitem[{{Dodelson}(2003)}]{2003moco.book.....D}
{Dodelson}, S. 2003, {Modern cosmology}

\bibitem[{{Gazta{\~n}aga} {et~al.}(2009{\natexlab{a}}){Gazta{\~n}aga},
  {Cabr{\'e}}, \& {Hui}}]{2009MNRAS.399.1663G}
{Gazta{\~n}aga}, E., {Cabr{\'e}}, A., \& {Hui}, L. 2009{\natexlab{a}}, \mnras,
  399, 1663

\bibitem[{{Gazta{\~n}aga} {et~al.}(2009{\natexlab{b}}){Gazta{\~n}aga},
  {Miquel}, \& {S{\'a}nchez}}]{2009PhRvL.103i1302G}
{Gazta{\~n}aga}, E., {Miquel}, R., \& {S{\'a}nchez}, E. 2009{\natexlab{b}},
  Physical Review Letters, 103, 091302

\bibitem[{{Jarosik} {et~al.}(2011){Jarosik}, {Bennett}, {Dunkley}, {Gold},
  {Greason}, {Halpern}, {Hill}, {Hinshaw}, {Kogut}, {Komatsu}, {Larson},
  {Limon}, {Meyer}, {Nolta}, {Odegard}, {Page}, {Smith}, {Spergel}, {Tucker},
  {Weiland}, {Wollack}, \& {Wright}}]{2011ApJS..192...14J}
{Jarosik}, N., {Bennett}, C.~L., {Dunkley}, J., {et~al.} 2011, \apjs, 192, 14

\bibitem[{{Jimenez} {et~al.}(2003){Jimenez}, {Verde}, {Treu}, \&
  {Stern}}]{2003ApJ...593..622J}
{Jimenez}, R., {Verde}, L., {Treu}, T., \& {Stern}, D. 2003, \apj, 593, 622

\bibitem[{{Kawamura} {et~al.}(2006){Kawamura}, {Nakamura}, {Ando}, {Seto},
  {Tsubono}, {Numata}, {Takahashi}, {Nagano}, {Ishikawa}, {Musha}, {Ueda},
  {Sato}, {Hosokawa}, {Agatsuma}, {Akutsu}, {Aoyanagi}, {Arai}, {Araya},
  {Asada}, {Aso}, {Chiba}, {Ebisuzaki}, {Eriguchi}, {Fujimoto}, {Fukushima},
  {Futamase}, {Ganzu}, {Harada}, {Hashimoto}, {Hayama}, {Hikida}, {Himemoto},
  {Hirabayashi}, {Hiramatsu}, {Ichiki}, {Ikegami}, {Inoue}, {Ioka},
  {Ishidoshiro}, {Itoh}, {Kamagasako}, {Kanda}, {Kawashima}, {Kirihara},
  {Kiuchi}, {Kobayashi}, {Kohri}, {Kojima}, {Kokeyama}, {Kozai}, {Kudoh},
  {Kunimori}, {Kuroda}, {Maeda}, {Matsuhara}, {Mino}, {Miyakawa}, {Miyoki},
  {Mizusawa}, {Morisawa}, {Mukohyama}, {Naito}, {Nakagawa}, {Nakamura},
  {Nakano}, {Nakao}, {Nishizawa}, {Niwa}, {Nozawa}, {Ohashi}, {Ohishi},
  {Ohkawa}, {Okutomi}, {Oohara}, {Sago}, {Saijo}, {Sakagami}, {Sakata},
  {Sasaki}, {Sato}, {Shibata}, {Shinkai}, {Somiya}, {Sotani}, {Sugiyama},
  {Tagoshi}, {Takahashi}, {Takahashi}, {Takahashi}, {Takano}, {Tanaka},
  {Taniguchi}, {Taruya}, {Tashiro}, {Tokunari}, {Tsujikawa}, {Tsunesada},
  {Yamamoto}, {Yamazaki}, {Yokoyama}, {Yoo}, {Yoshida}, \&
  {Yoshino}}]{2006CQGra..23S.125K}
{Kawamura}, S., {Nakamura}, T., {Ando}, M., {et~al.} 2006, Classical and
  Quantum Gravity, 23, S125

\bibitem[{{Keppel} \& {Ajith}(2010)}]{2010PhRvD..82l2001K}
{Keppel}, D., \& {Ajith}, P. 2010, \prd, 82, 122001

\bibitem[{{Ma} \& {Zhang}(2011)}]{2011ApJ...730...74M}
{Ma}, C., \& {Zhang}, T.-J. 2011, \apj, 730, 74

\bibitem[{{Meng} {et~al.}(2015){Meng}, {Wang}, {Li}, \&
  {Zhang}}]{2015arXiv150702517M}
{Meng}, X.-L., {Wang}, X., {Li}, S.-Y., \& {Zhang}, T.-J. 2015, ArXiv e-prints,
  arXiv:1507.02517

\bibitem[{{Moresco}(2015)}]{2015MNRAS.450L..16M}
{Moresco}, M. 2015, \mnras, 450, L16

\bibitem[{{Moresco} {et~al.}(2012){Moresco}, {Verde}, {Pozzetti}, {Jimenez}, \&
  {Cimatti}}]{2012JCAP...07..053M}
{Moresco}, M., {Verde}, L., {Pozzetti}, L., {Jimenez}, R., \& {Cimatti}, A.
  2012, \jcap, 7, 053

\bibitem[{{Moresco} {et~al.}(2016){Moresco}, {Pozzetti}, {Cimatti}, {Jimenez},
  {Maraston}, {Verde}, {Thomas}, {Citro}, {Tojeiro}, \&
  {Wilkinson}}]{2016JCAP...05..014M}
{Moresco}, M., {Pozzetti}, L., {Cimatti}, A., {et~al.} 2016, \jcap, 5, 014

\bibitem[{{Nishizawa} {et~al.}(2011){Nishizawa}, {Taruya}, \&
  {Saito}}]{2011PhRvD..83h4045N}
{Nishizawa}, A., {Taruya}, A., \& {Saito}, S. 2011, \prd, 83, 084045

\bibitem[{{Riess} {et~al.}(2009){Riess}, {Macri}, {Casertano}, {Sosey},
  {Lampeitl}, {Ferguson}, {Filippenko}, {Jha}, {Li}, {Chornock}, \&
  {Sarkar}}]{2009ApJ...699..539R}
{Riess}, A.~G., {Macri}, L., {Casertano}, S., {et~al.} 2009, \apj, 699, 539

\bibitem[{{Samushia} {et~al.}(2013){Samushia}, {Reid}, {White}, {Percival},
  {Cuesta}, {Lombriser}, {Manera}, {Nichol}, {Schneider}, {Bizyaev},
  {Brewington}, {Malanushenko}, {Malanushenko}, {Oravetz}, {Pan}, {Simmons},
  {Shelden}, {Snedden}, {Tinker}, {Weaver}, {York}, \&
  {Zhao}}]{2013MNRAS.429.1514S}
{Samushia}, L., {Reid}, B.~A., {White}, M., {et~al.} 2013, \mnras, 429, 1514

\bibitem[{{Sasaki}(1987)}]{1987MNRAS.228..653S}
{Sasaki}, M. 1987, \mnras, 228, 653

\bibitem[{{Simon} {et~al.}(2005){Simon}, {Verde}, \&
  {Jimenez}}]{2005PhRvD..71l3001S}
{Simon}, J., {Verde}, L., \& {Jimenez}, R. 2005, \prd, 71, 123001

\bibitem[{{Spergel} {et~al.}(2007){Spergel}, {Bean}, {Dor{\'e}}, {Nolta},
  {Bennett}, {Dunkley}, {Hinshaw}, {Jarosik}, {Komatsu}, {Page}, {Peiris},
  {Verde}, {Halpern}, {Hill}, {Kogut}, {Limon}, {Meyer}, {Odegard}, {Tucker},
  {Weiland}, {Wollack}, \& {Wright}}]{2007ApJS..170..377S}
{Spergel}, D.~N., {Bean}, R., {Dor{\'e}}, O., {et~al.} 2007, \apjs, 170, 377

\bibitem[{{Stern} {et~al.}(2010){Stern}, {Jimenez}, {Verde}, {Kamionkowski}, \&
  {Stanford}}]{2010JCAP...02..008S}
{Stern}, D., {Jimenez}, R., {Verde}, L., {Kamionkowski}, M., \& {Stanford},
  S.~A. 2010, \jcap, 2, 008

\bibitem[{{Xu} {et~al.}(2013){Xu}, {Cuesta}, {Padmanabhan}, {Eisenstein}, \&
  {McBride}}]{2013MNRAS.431.2834X}
{Xu}, X., {Cuesta}, A.~J., {Padmanabhan}, N., {Eisenstein}, D.~J., \&
  {McBride}, C.~K. 2013, \mnras, 431, 2834

\bibitem[{{Yuan} \& {Zhang}(2015)}]{2015JCAP...02..025Y}
{Yuan}, S., \& {Zhang}, T.-J. 2015, \jcap, 2, 025

\bibitem[{{Zhang} {et~al.}(2014){Zhang}, {Zhang}, {Yuan}, {Liu}, {Zhang}, \&
  {Sun}}]{2014RAA....14.1221Z}
{Zhang}, C., {Zhang}, H., {Yuan}, S., {et~al.} 2014, Research in Astronomy and
  Astrophysics, 14, 1221

\end{thebibliography}

\end{document}